\documentstyle[11pt]{article}

\newcommand{\eq}{\begin{equation}}
\newcommand{\en}{\end{equation}}
\newcommand{\eqn}{\begin{eqnarray}}
\newcommand{\enn}{\end{eqnarray}}

\newcommand{\beq}{\begin{equation}}
\newcommand{\eeq}{\end{equation}}

\begin{document}

\title{\vspace*{-1.5in} {\ {\small \leftline{USC-99/HEP-B4 \hfill  hep-th/9910091 }%
\vspace{-10pt} \leftline{CITUSC/99-002 } } \vskip1in } Non-Commutative
Geometry\\
on a Discrete Periodic Lattice\\
and Gauge Theory\thanks{%
Research partially supported by the U.S. Department of Energy under grant
number DE-FG03-84ER40168.} \bigskip\medskip }
\author{{\bf Itzhak Bars and Djordje Minic\thanks{%
e-mail: bars@physics.usc.edu, minic@physics.usc.edu} \bigskip} \\
Caltech-USC Center for Theoretical Physics\\
\smallskip and\\
\smallskip Department of Physics and Astronomy\\
University of Southern California\\
Los Angeles, CA 90089-0484}
\date{}
\maketitle

\begin{abstract}
We discuss the quantum mechanics of a particle in a magnetic field when its
position $x^{\mu }$ is restricted to a periodic lattice, while its momentum $%
p^{\mu }$ is restricted to a periodic dual lattice. Through these
considerations we define non-commutative geometry on the lattice. This leads
to a deformation of the algebra of functions on the lattice, such that their
product involves a ``diamond'' product, which becomes the star product in
the continuum limit. We apply these results to construct non-commutative
U(1) and U$\left( M\right) $ gauge theories, and show that they are
equivalent to a pure U$\left( NM\right) $ matrix theory, where $N^{2}$ is
the number of lattice points. \newpage
\end{abstract}

\section{Introduction and results}

Recently non-commutative geometry has found applications in string and
M-theory in the $B$-field background \cite{nonc}\cite{seibwitnonc}. The
non-commutative geometry in question is described by a deformation of the
ordinary algebra of functions $f\left( x\right) g\left( x\right) $ on R$^{d}$
into a non-commutative albeit associative algebra, with a star product \cite
{seibwitnonc} 
\begin{equation}
f\left( x\right) \ast g\left( x\right) =\left. \exp \left( \frac{i}{2}\theta
^{\mu \nu }\frac{\partial }{\partial x^{\mu }}\frac{\partial }{\partial
y^{\nu }}\right) \,f\left( x\right) g\left( y\right) \right| _{y=x}.
\label{starp}
\end{equation}
$\theta ^{\mu \nu }$ is given in terms of a constant background B-field that
has even rank $d$ 
\begin{equation}
\theta ^{\mu \nu }=-\left( 2\pi \alpha ^{\prime }\right) ^{2}\left( \frac{1}{%
g+2\pi \alpha ^{\prime }B}B\frac{1}{g-2\pi \alpha ^{\prime }B}\right) ^{\mu
\nu }.  \label{theta}
\end{equation}
In the limit $\alpha ^{\prime }\rightarrow 0$ and $g_{\mu \nu }\sim \left(
\alpha ^{\prime }\right) ^{2},$ string theory is correctly represented by
non-commutative gauge theory, with $\theta ^{\mu \nu }=\left( B^{-1}\right)
^{\mu \nu }.$ Effectively this is the large $B$ limit. The indices $\mu $
label a Euclidean space\footnote{%
In the string theory derivation, these dimensions correspond to the
Euclidean dimensions of a D-brane along which the string background field $%
B_{\mu \nu }$ does not vanish. The field $B_{\mu \nu }$ can also be thought
of as a constant magnetic field with a potential $A_{\mu }=\frac{1}{2}x^{\nu
}B_{\nu \mu }$ interacting with charged points $x^{\mu }\left( \tau \right) $
at the end of a string. However, taken as an independent starting point,
there does not seem to be any problem in allowing one of the dimensions to
be timelike. Our discussion does not change if the space is purely Euclidean
or Minkowski.} $\mu =1,\cdots ,d.$ The star product is related to the Moyal
bracket \cite{(7)}\cite{(8)}\cite{(11)}\cite{star}. When this product is
used instead of the ordinary product of functions in a gauge theory, the
resulting non-commutative gauge theory represents string theory in a large $%
B_{\mu \nu }$ limit, including the non-perturbative effects of the
background B-field \cite{nonc}\cite{seibwitnonc}.

In order to further analyze non-commutative gauge theory, a cutoff version
would be useful. With this in mind we define non-commutative gauge theory on
a discrete periodic lattice that has two parameters: the periodicity
characterized by a length $L$ and the lattice spacing $a$. The ratio of
these is the number of steps $n=L/a$ in each direction labelled by $\mu .$
In effect, this lattice is the d-dimensional discretized torus $T^{d}$ in
d-dimensions, with $n$ steps in every direction, which we will denote by $%
\left( T_{n}\right) ^{d}.$ There are altogether $n^{d}$ lattice points on
the discrete torus. A less uniform lattice would have different number of
steps in the various directions $\mu ,$ such that the total number of
lattice points would be $\prod n_{\mu }$, instead of $n^{d}.$ In most of the
paper we will concentrate on the uniform lattice for simplicity, but we will
also discuss some interesting aspects of a non-uniform lattice in which the
number of lattice points is not the same in every direction, but are taken
equal in pairs, such that $n_{1}$ for both $\mu =1,2,$ and $n_{2}$ for both $%
\mu =3,4,$ etc. By identifying 
\begin{equation}
N=n_{1}n_{2}\cdots n_{d/2},
\end{equation}
($d$ is even) we see that the positions $x^{\mu }$ live on the $N^{2}$
points of the periodic lattice $\left( T_{n}\right) ^{d}$. The gauge fields $%
A_{\mu }\left( x\right) $ or other functions on the lattice are defined only
these $N^{2}$ spacetime points.

We then construct a ``diamond product'' which is a lattice version of the
star product. We will be guided by a previous construction that introduced
the discrete Moyal bracket \cite{bars} as a cutoff version of the Moyal
bracket with a different application in mind \cite{(1)}-\cite{hoppe}. The
first step is to provide an explicit map $\hat{\Delta}_{I}^{J}\left(
x\right) $ from the $N^{2}$ lattice points $x^{\mu }$ to a $N\times N$
matrix that has $N^{2}$ entries. Then any function $f\left( x^{\mu }\right) $
defined on the $N^{2}$ lattice points can be rewritten in terms of a matrix $%
\hat{f}$ with $x$-independent matrix elements $\hat{f}_{I}^{J},$ $%
I,J=1,2,\cdots ,N,$ as follows (matrices are denoted by the hat symbol) 
\begin{equation}
f\left( x\right) =\frac{1}{N}Tr\left( \hat{\Delta}\left( x\right) \hat{f}%
\right) ,\quad \hat{f}_{I}^{J}=\sum_{x\in \left( T_{n}\right) ^{d}}\hat{%
\Delta}_{I}^{J}\left( x\right) f\left( x\right) .  \label{map}
\end{equation}

The properties of the map $\hat{\Delta}_{I}^{J}\left( x\right) $ are
obtained by studying the quantum mechanics of particles in a constant
magnetic field $B_{\mu \nu }$, such that the particle positions $x^{\mu }$
are at the $n^{d}$ lattice points on $\left( T_{n}\right) ^{d},$ while their
momenta $p_{\mu }$ ($\partial /\partial x^{\mu }$ in continuum) are on $%
n^{d} $ points on the dual lattice. The dual lattice $\left( \tilde{T}%
_{n}\right) ^{d}$ is similar to $\left( T_{n}\right) ^{d}$ but its lattice
spacing is measured in terms of momentum units. Then the map is given by 
\begin{equation}
\hat{\Delta}_{I}^{J}\left( x\right) =\frac{1}{N}\sum_{p^{\mu }\in \left( 
\tilde{T}_{n}\right) ^{d}}e^{-ip\cdot x}\left[ \exp \left( ip\cdot X\right) %
\right] _{I}^{J}\,\,
\end{equation}
where $\left[ \exp \left( ip\cdot X\right) \right] _{I}^{J}\,\,$ is a matrix
that will be given explicitly. Roughly, this map is the matrix elements of a
delta function $\delta ^{\left( d\right) }\left( X-x\right) $ with $X^{\mu }$
non-commutative operators and $x^{\mu }$ defined only on the periodic
lattice. The map contains all the information about non-commutative geometry
on the periodic lattice. Using this map and the definitions in (\ref{map}),
the diamond product is constructed as follows \cite{bars} 
\begin{eqnarray}
f\left( x\right) \diamond g\left( x\right) &=&\frac{1}{N}Tr\left( \hat{\Delta%
}\left( x\right) \hat{f}\hat{g}\right) , \\
&=&\frac{1}{N}\sum_{y,z\in \left( T_{n}\right) ^{d}}f\left( y\right) g\left(
z\right) \,\exp \left( 2iB^{\mu \nu }\left( x_{\mu }-y_{\mu }\right) \left(
x_{\nu }-z_{\nu }\right) \right) .  \nonumber
\end{eqnarray}
It is physically interesting to note that the sums in the diamond product
are weighted by exponentials of the flux that passes through the area
defined by the three lattice points $x,y,z$. We will show that the diamond
product reduces to the star product (\ref{starp}) in the continuum limit. In
this way the diamond (or star) product is explicitly related to ordinary
matrix product $\left( \hat{f}\hat{g}\right) _{I}^{J}.$

Using this formulation we show that the non-commutative U$\left( 1\right) $
gauge theory on the periodic lattice can be rewritten as a U$\left( N\right) 
$ pure matrix theory where all spacetime positions $x^{\mu }$ have been
converted to matrix elements by using the map. The non-Abelian U$\left(
M\right) $ non-commutative gauge theory on the lattice can also be discussed
in the same non-commutative formalism by generalizing to a U$\left(
NM\right) $ matrix theory.

The $U\left( M\right) $ non-commutative gauge theory action on the periodic
lattice is constructed by using the diamond product $A_{\mu }\left( x\right)
\diamond A_{\nu }\left( x\right) $ whenever gauge fields need to be
multiplied with each other, and by substituting the derivative $\partial
_{\mu }A_{\nu }\left( x\right) $ by a suitable lattice version, but
otherwise keeping the same general form of the Yang-Mills action. By using
the map $\hat{\Delta}_{I}^{J}\left( x\right) $ the U$\left( M\right) $
lattice action is rewritten in the following pure matrix version 
\begin{equation}
S=\frac{1}{4N^{2}}\sum_{x\in \left( T_{n}\right) ^{d}}\left( F_{\mu \nu
}\left( x\right) \right) _{a}^{a^{\prime }}\diamond \left( F_{\mu \nu
}\left( x\right) \right) _{a^{\prime }}^{a}=-\frac{1}{4}Tr\left( \left[
a_{\mu },a_{\nu }\right] ^{2}\right) ,  \label{straa}
\end{equation}
where the $x^{\mu }$-independent $a_{\mu }$ is an $NM\times NM$ matrix
related to the $M\times M$ gauge field $\left( A_{\mu }\left( x\right)
\right) _{a}^{a^{\prime }}$ in a way that will be indicated.

It is possible to interpret the non-Abelian U$\left( M\right) $ theory in $d$
-dimensions as an Abelian U$\left( 1\right) $ theory in $d+2$ dimensions.
This comes about by considering a non-uniform lattice as described above.
Then the U$\left( 1\right) $ Abelian theory in $d$ dimensions is described
by a U$\left( N\right) $ pure matrix theory (\ref{straa}) with $%
N=n_{1}n_{2}\cdots n_{d/2},$ whereas the non-Abelian U$\left( M\right) $
theory in $d$ dimensions can be regarded as a U$\left( 1\right) $ theory
with two more non-commutative discretized dimensions, with lattice steps $%
n_{\left( d+2\right) /2}\equiv M,$ so that $NM=n_{1}n_{2}\cdots
n_{d/2}n_{\left( d+2\right) /2}.$ Thus, in the U$\left( NM\right) $ matrix
theory (\ref{straa}), $N=n_{1}n_{2}\cdots ,n_{d/2}$ relates to space and $%
M\equiv n_{\left( d+2\right) /2}$ relates to two more non-commutative
discrete dimensions that replace the internal space.

The form of the action (\ref{straa}) could be related to the reduced models
of gauge theories \cite{(12)}, or more precisely, to the fully reduced
Matrix theory version written in the form $Tr\left( \left[ X_{\mu },X_{\nu } %
\right] ^{2}\right) $ \cite{(6)}. However, in the present version, the
physical meaning of the matrix is quite different. Namely, the space-time
interpretation is obtained via the map $\hat{\Delta}_{I}^{J}\left( x\right) $
related to the factor $N=n_{1}n_{2}\cdots ,n_{d/2},$ and the internal
symmetry information is in the factor $M,$ which are different than the
spacetime/internal symmetry interpretation of the reduced models. Thus the
existing computational technology of reduced models and matrix models could
be adapted to the current problem provided one takes care of the physical
interpretation via the map (\ref{map}) and the meaning of $N,M$. Some recent
computations in \cite{ikkt}\cite{iikk} also seem to be related to our
observations, but with a somewhat different spacetime interpretation.

The organization of this paper is as follows: First we discuss the quantum
mechanics of particles in a lattice in a magnetic field and show how to
derive non-commutative geometry on the lattice from such considerations.
This leads directly to an explicit expression for the map $\hat{\Delta}
_{I}^{J}\left( x\right) .$ We apply these results to the non-commutative
U(1) and U$\left( M\right) $ gauge theories on the lattice, and show that
they are equivalent to a pure U$\left( NM\right) $ matrix theory, as in (\ref
{straa}).

The larger project of studying non-commutative gauge theories in this cutoff
version should be worthwhile, but it is not pursued in the current paper.

\section{Non-commutative geometry on the lattice}

It is well known that the quantum mechanics of a particle in a constant
magnetic field $B_{\mu \nu }$ produces non-commutative momenta \cite{zak} $%
\left[ K_{\mu },K_{\nu }\right] =iB_{\mu \nu }.$ In order to map this
problem to the string theory setting we define ``coordinates'' $X_{\mu
}=\left( B^{-1}\right) _{\mu \nu }K^{\nu }$ which satisfy the commutation
rules of non-commutative geometry \cite{nonc}\cite{seibwitnonc} 
\begin{equation}
\left[ X^{\mu },X^{\nu }\right] =i\left( B^{-1}\right) ^{\mu \nu }\equiv
i\theta ^{\mu \nu }.
\end{equation}
For simplicity, we begin the discussion of the lattice version of this setup
for the special form of $B^{\mu \nu }$ that is block diagonal, with 2$\times
2$ blocks along the diagonal, each of them proportional to the Pauli matrix $%
i\sigma _{2}$ with various proportionality constants, and zero entries
otherwise (it is always possible to rotate $B^{\mu \nu }$ into such a
basis). At the end we generalize to an arbitrary form of $B^{\mu \nu }.$ For
the special form of $\theta ^{\mu \nu }$ non-commutativity occurs in pairs
of coordinates 
\begin{equation}
\left[ X^{1},X^{2}\right] =\frac{i}{B_{12}},\quad \left[ X^{3},X^{4}\right] =%
\frac{i}{B_{34}},\quad \cdots  \label{x1x2}
\end{equation}
There does not seem to be anything special about a distinction between
timelike or spacelike coordinates since all signs may be absorbed into a
redefinition of $B_{\mu \nu }$. We will first discuss the pair $\left(
X^{1},X^{2}\right) $ and later include all the $X^{\mu }$. We will then
follow the construction of the 2D diamond product in \cite{bars} whose
discussion we generalize to higher dimensions.

\subsection{Two-torus}

Since $X^{1},X^{2}$ do not commute, they cannot be diagonalized
simultaneously. Consider diagonalizing $X^{1}.$ In the continuum the
eigenvalues are on the real line. Consider a periodic lattice, with period $%
L $ and lattice spacing $a$ in the $X^{1}$ direction. The eigenstates of $%
X^{1} $ are labelled as $|j_{1}>,$ $j_{1}=0,1,\cdots ,n-1,$ with $n=L/a,$
and the eigenvalues of $X^{1}$ are restricted to the discrete set $%
x^{1}=aj_{1}$. Furthermore there is a periodicity condition 
\begin{equation}
X^{1}|j_{1}>=aj_{1}|j_{1}>,\quad |j_{1}+n>=|j_{1}>,  \label{eigen1}
\end{equation}
therefore the eigenvalues $x^{1}$ take discrete values on the circle of
perimeter $L$ 
\begin{equation}
x^{1}=a\left( j_{1}{\rm mod\,}n\right) .
\end{equation}
According to (\ref{x1x2}) the operator $B_{12}X_{2}$ acts like infinitesimal
translations on the eigenspace of $X^{1}.$ On the lattice only finite
translations make sense. Taking the commutation rules (\ref{x1x2}) into
account, the translation operator by one lattice unit is $\exp \left(
iaB_{12}X^{2}\right) $ 
\begin{equation}
<j_{1}\,|\exp \left( iaX^{2}B_{12}\right) =<j_{1}+1|.
\end{equation}
Its matrix elements take the form 
\begin{equation}
\hat{g}_{j_{1}}^{\,\,j_{1}^{\prime }}=<j_{1}|\exp \left(
iaX^{2}B_{12}\right) \,|j_{1}^{\prime }>=\delta _{\left( 1+j_{1}\right) {\rm %
mod\,}n}^{\,\,j_{1}^{\prime }{\rm mod\,}n}
\end{equation}
Including the periodicity condition, $\hat{g}_{j_{1}}^{\,\,j_{1}^{\prime }}$
becomes the well known circular matrix that has also a non-trivial entry in
location $\hat{g}_{n-1}^{\,\,0}=1$%
\begin{equation}
\hat{g}=\left( 
\begin{array}{ccccc}
0 & 1 & 0 & \cdots & 0 \\ 
0 & 0 & 1 & \ddots & \vdots \\ 
\vdots & 0 & 0 & \ddots & 0 \\ 
0 & \vdots & \ddots & \ddots & 1 \\ 
1 & 0 & \cdots & 0 & 0
\end{array}
\right) .
\end{equation}
$k_{1}$ units of translation along $X^{1}$ is obtained by taking $k_{1}$
powers of $\hat{g}$%
\begin{equation}
\exp \left( ik_{1}aX^{2}B_{12}\right) \rightarrow \left( \hat{g}%
^{k_{1}}\right) _{j_{1}}^{\,\,j_{1}^{\prime }}.
\end{equation}
Due to periodicity, $n$ units of translation must give the same state.
Indeed this is reflected in the property of the circular matrix $\hat{g}$%
\begin{equation}
\hat{g}^{n}=1.
\end{equation}

Similarly we consider diagonalizing $X^{2}$ on a periodic lattice with
periodicity $L$ and lattice spacing $a$ in the $X^{2}$ direction such that $%
an=L$. The eigenstates $|j_{2}>$ are associated with the eigenvalues $%
a\left( j_{2}{\rm mod\,}n\right) .$ In the eigenspace of $X^{2}$ the
translation operator by one unit is $\exp \left( -iaX^{1}B_{12}\right) $ and
it has similar properties to $\hat{g}$. However, acting on the eigenspace of 
$X^{1}$ defined in (\ref{eigen1}), this operator is a diagonal matrix 
\begin{equation}
\hat{h}_{j_{1}}^{\,\,j_{1}^{\prime }}=<j_{1}|\exp \left(
-iaX^{1}B_{12}\right) \,|j_{1}^{\prime }>=\delta _{j_{1}}^{\,\,j_{1}^{\prime
}}\,e^{-i\left( j_{1}{\rm mod\,}n\right) a^{2}B_{12}}.
\end{equation}
Taking into account the periodicity of the lattice in the $X^{2}$ direction, 
$n$ powers of $\hat{h}$ should be the identity operator for any state. This
requires 
\begin{equation}
a^{2}B_{12}=\frac{2\pi b_{12}}{n}
\end{equation}
where $b_{12}$ is an integer. Therefore the magnetic flux $a^{2}B_{12}$
passing through a lattice unit surface $a^{2}$ in the 1-2 plane is quantized
as $b_{12}$ units of $2\pi /n$.

It is convenient to define $\omega $ as the $n$-th root of the identity 
\begin{equation}
\omega =\exp \left( -ia^{2}B_{12}\right) =e^{-i\frac{2\pi b_{12}}{n}},\quad
\omega ^{n}=1.
\end{equation}
The matrix elements of $\hat{h}$ can then be written in the form 
\begin{equation}
\hat{h}=\left( 
\begin{array}{ccccc}
1 & 0 & 0 & \cdots & 0 \\ 
0 & \omega & 0 & \cdots & 0 \\ 
0 & 0 & \omega ^{2} & \ddots & \vdots \\ 
\vdots & \vdots & \ddots & \ddots & 0 \\ 
0 & 0 & \cdots & 0 & \omega ^{n-1}
\end{array}
\right) .
\end{equation}
If one diagonalizes the matrix $\hat{g},$ the result must be the matrix $%
\hat{h}$ since the roles of $X^{1},X^{2}$ can be reversed. Indeed, one can
find the explicit unitary transformation 
\begin{equation}
\hat{g}=\hat{U}\hat{h}\hat{U}^{\dagger },\quad \hat{U}_{j}^{\,\,j^{\prime }}=%
\frac{1}{\sqrt{n}}\omega ^{jj^{\prime }}.  \label{U}
\end{equation}
The unitary matrix $\hat{U}$ also satisfies the periodicity property under $%
j\rightarrow j+n$ thanks to the fact that $\omega $ is the $n$-th root of
unity. The commutation property of these matrices is well known 
\begin{equation}
\hat{g}\hat{h}=\hat{h}\hat{g}\omega .\quad
\end{equation}
They follow from the non-commutative properties of the coordinates $\left[
X^{1},X^{2}\right] =i/B_{12}$ by using $\exp \left( \beta X^{2}\right) \exp
\left( \alpha X^{1}\right) \,=\exp \left( \alpha X^{1}\right) \,\exp \left(
\beta X^{2}\right) \exp \left[ \beta X^{2},\alpha X^{1}\right] $. $%
\allowbreak $Thus the matrices $\hat{g},\hat{h}$ capture the essence of the
non-commutative geometry on the lattice.

On the entire quantum space, whether $X^{1}$ or $X^{2}$ is diagonal, the
only operators that are meaningful are all the possible translations given
by $\exp \left( ia\left( k^{1}X^{2}-k^{2}X^{1}\right) B_{12}\right) $ with $%
k_{1},k_{2}$ integers modulo $n.$ Their matrix elements are given by 
\begin{equation}
<j_{1}|\exp \left( ia\left( k^{1}X^{2}-k^{2}X^{1}\right) B_{12}\right)
|j_{1}^{\prime }>=\omega ^{k_{1}k_{2}/2}\left( \hat{h}^{k_{2}}\hat{g}%
^{k_{1}}\right) _{j_{1}}^{j_{1}^{\prime }}\equiv \left( \hat{v}%
_{k_{1}k_{2}}\right) _{j_{1}}^{j_{1}^{\prime }}  \label{lk1k2}
\end{equation}
where we have used the formula $\exp \left( A+B\right) =\exp A\,\exp B\,\exp
\left( -\left[ A,B\right] /2\right) $ on the left hand side and then
evaluated the matrix elements.

It is useful to define a momentum lattice given by $p_{\mu }=ak^{\nu }B_{\nu
\mu }$ where the lattice distance is measured by $aB_{12}$ and the integers $%
k_{1},k_{2}$ are defined modulo $n.$ 
\begin{equation}
p_{1}=-aB_{12}\left( k_{2}{\rm mod\,}n\right) ,\quad p_{2}=aB_{12}\left(
k_{1}{\rm mod\,}n\right)
\end{equation}
This lattice is the {\it dual lattice} to the position lattice, its steps
are measured in units of momentum. Then the full set of $n^{2}$ translation
operators take a more suggestive form of a plane wave operator, or ``vertex
operator'', whose matrix elements are $\left( \hat{v}_{p^{\mu }}\right)
_{j_{1}}^{j_{1}^{\prime }}$%
\begin{equation}
\exp \left( ip_{\mu }X^{\mu }\right) \rightarrow \hat{v}_{p}=\hat{h}^{k_{2}}%
\hat{g}^{k_{1}}\omega ^{k_{1}k_{2}/2}.
\end{equation}
These translations are the only meaningful operators that need to be
considered for the quantum mechanics of the particle on the non-commutative
discrete torus. They have the well known property that under matrix
multiplication they form a group algebra 
\begin{eqnarray}
\hat{v}_{p}\hat{v}_{p^{\prime }} &=&\,\hat{v}_{p+p^{\prime }}\,\,\omega
^{\left( \frac{1}{2}\varepsilon ^{\mu \nu }k_{\mu }k_{\nu }^{\prime }\right)
}  \label{galg} \\
&=&\hat{v}_{p+p^{\prime }}\,\exp \left( -i\frac{2\pi b_{12}}{2n}\left(
k_{1}k_{2}^{\prime }-k_{2}k_{1}^{\prime }\right) \right)
\end{eqnarray}
which is derived by using $\hat{g}^{a}\hat{h}^{b}=\hat{h}^{b}\hat{g}%
^{a}\omega ^{ab}.$

One final remark is in order: 
To avoid confusion, one should not think of $h$, $g$ as being obtained by 
exponentiating matrices $X^1$, $X^2$ that are $n \times n$ matrices. This is {\it not} how 
we presented them. Rather, we have evaluated the matrix elements of 
exponentials of the {\it operators} $X^1$, $X^2$ and obtained finite $n \times n$ 
matrices, because we used only a discrete set of states that represent the 
$n$ 
lattice points. We argued that only exponentials of operators $X^1$, $X^2$ that 
correspond to finite lattice translations are needed to discuss the lattice. 
These exponentials clearly are finite $n \times n$  matrices on the lattice, as we have 
seen. For discussing the lattice, only discrete powers of the same 
exponentials are used, while other functions of the operators  $X^1$, $X^2$
are never 
needed. Thus, while $X^1$, $X^2$ are opearators acting on an infinite Hilbert space, 
only a finite set of that space comes into play thanks to the fact that only 
the exponentials of $X^1$, $X^2$ enter in the lattice theory. The infinite space 
becomes relevant when the lattice spacing goes to zero or $n$ goes to infinity. 
If one wishes, one may {\it define} $n \times n$ matrices 
$\tilde{X^{1}}$, $\tilde{X^{2}}$ as 
the logarithms of the matrices  $h$, $g$ but these should not be identified with 
the operators $X^1$, $X^2$. Obviously the commutation rules in eq.(9) are true for 
the operators $X^1$, $X^2$ but not true for the matrices $\tilde{X^{1}}$, 
$\tilde{X^{2}}$. As $n$ 
goes to infinity $\tilde{X^{1}}$, $\tilde{X^{2}}$ would approach the matrix elements of 
the operators $X^1$, $X^2$.

\subsection{$d$- dimensions}

Now we generalize the previous section to $d$-dimensions. Consider any other
pair in the set of non-commuting operators, such as $X^{3},X^{4}.$ The story
is the same as in the previous section. The eigenspace of the operator $%
X^{3} $ is labelled by $|j_{3}{\rm mod\,}n>,$ and the eigenvalues are $%
x^{3}=a(j_{3}{\rm mod\,}n).$ The set of all operators that need to be
considered are $\exp \left( ip_{3}X^{3}+ip_{4}X^{4}\right) $ with 
\begin{equation}
p_{3}=-aB_{34}\left( k^{3}{\rm mod\,}n\right) ,\quad p_{4}=aB_{34}\left(
k^{4}{\rm mod\,}n\right) ,
\end{equation}
and with a quantization rule for the flux 
\begin{equation}
a^{2}B_{34}=\frac{2\pi b_{34}}{n},\quad b_{34}={\rm integer},
\end{equation}
that leads to a phase $\omega _{34}$ 
\begin{equation}
\omega _{34}=\exp \left( -ia^{2}B_{34}\right) =e^{-i\frac{2\pi b_{34}}{n}%
},\quad \left( \omega _{34}\right) ^{n}=1.
\end{equation}
The corresponding translation matrices $\hat{h}_{34},\hat{g}_{34}$ satisfy $%
\hat{g}_{34}\hat{h}_{34}=\hat{h}_{34}\hat{g}_{34}\omega _{34}$ and they lead
to the group algebra (\ref{galg}) with $\omega _{34},k_{3},k_{4}$ inserted
instead of $\omega ,k_{1},k_{2}.$

The combined non-commutative geometry for all the operators can be treated
by taking a direct product of the eigenspaces of $X^{1},X^{3},X^{5},\cdots
,X^{d-1}$%
\begin{equation}
|j_{1},j_{3},\cdots ,j_{d-1}>,\quad x_{2i-1}=a\left( j_{2i-1}{\rm mod\,}%
n\right) .  \label{tim1}
\end{equation}
The remaining operators $X_{2},X_{4},\cdots ,X_{d}$ cannot be simultaneously
diagonalized with the above. But in the space in which they are diagonal
(obtained by applying the transformation $U$ in (\ref{U})) these $X_{2i}$
have eigenvalues that are similar to those above $x_{2i}=a\left( j_{2i}{\rm %
mod\,}n\right) .$

The flux is quantized because of the periodicity of the lattice 
\begin{eqnarray}
a^{2}B_{\mu \nu } &=&\frac{2\pi b_{\mu \nu }}{n},\quad b_{\mu \nu }={\rm %
integer} \\
\omega _{\mu \nu } &=&\exp \left( -ia^{2}B_{\mu \nu }\right) ,\quad \left(
\omega _{\mu \nu }\right) ^{n}=1,
\end{eqnarray}
and the momentum lattice is defined by 
\begin{equation}
p_{\mu }=ak^{\nu }B_{\nu \mu }=k^{\nu }\frac{2\pi b_{\nu \mu }}{an},\quad
p_{\mu }\in \left( \tilde{T}_{n}\right) ^{d}
\end{equation}
The set of all possible lattice translations $\exp \left( i\sum_{\mu
=1}^{d}p_{\mu }X^{\mu }\right) ,$ which is similar to a ``vertex operator''
in string theory, has the matrix elements 
\begin{eqnarray}
&<&j_{1},j_{3},\cdots ,j_{d-1}|\exp \left( i\sum_{\mu =1}^{d}p_{\mu }X^{\mu
}\right) |j_{1}^{\prime },j_{3}^{\prime },\cdots ,j_{d-1}^{\prime }> \\
&\equiv &\left( \hat{V}_{p}\right) _{j_{1}j_{3}\cdots
j_{d-1}}^{j_{1}^{\prime }j_{3}^{\prime }\cdots j_{d-1}^{\prime }}=\left( 
\hat{v}_{p_{1}p_{2}}\right) _{j_{1}}^{j_{1}^{\prime }}\cdots \left( \hat{v}%
_{p_{d-1}p_{d}}\right) _{j_{d-1}}^{j_{d-1}^{\prime }}
\end{eqnarray}
In matrix notation, the set of all translation operators takes the direct
product form 
\begin{equation}
\hat{V}_{p}=\hat{v}_{p_{1}p_{2}}\otimes \hat{v}_{p_{3}p_{4}}\otimes \cdots
\otimes \hat{v}_{p_{d-1}p_{d}}.  \label{Tp}
\end{equation}
The matrices $\hat{V}_{p}$ satisfy the group algebra 
\begin{eqnarray}
\hat{V}_{p}\hat{V}_{p^{\prime }} &=&\hat{V}_{p+p^{\prime }}\,\exp \left( -i%
\frac{\pi b_{\mu \nu }}{n}k_{\mu }k_{\nu }^{\prime }\right)  \label{galgeb}
\\
&=&\hat{V}_{p+p^{\prime }}\,\exp \left( -\frac{i}{2}\theta ^{\mu \nu }p_{\mu
}p_{\nu }^{\prime }\right)
\end{eqnarray}
which follows from (\ref{galg}). Under tracing one gets a Kronecker delta
function 
\begin{equation}
Tr\hat{V}_{p}=N\,\delta _{p_{\mu },0},\quad Tr\left( \hat{V}_{p}\hat{V}%
_{p^{\prime }}\right) =N\,\delta _{p_{\mu },-p_{\mu }^{\prime }}.
\label{trVp}
\end{equation}

The relation (\ref{galgeb}) looks formally the same as the continuum, but in
the present case it takes into account the momentum lattice $\left( \tilde{T}%
_{n}\right) ^{d}$ by having discrete momenta $p_{\mu }$, and the position
lattice $\left( T_{n}\right) ^{d}$ by taking discrete eigenvalues $x_{\mu }$%
. Only half of the $x_{\mu }$ label the matrix elements of the matrices $%
\left( \hat{V}_{p}\right) _{J}^{J^{\prime }}$ where $J$ is a label for the
direct product space $J=\left( j_{1},j_{3},j_{5},\cdots ,j_{d-1}\right) .$
For the more general lattice the rank of these matrices is $%
N=n_{1}n_{2}\cdots n_{d/2}.$ Both the position and momentum lattices are
periodic and this is manifest in the expression (\ref{Tp}) for $\hat{V}_{p}$.

Although this result was derived by taking a block diagonal $\theta _{\mu
\nu },$ it is easy to generalize. The final result (\ref{galgeb}) is valid
for the general quantized antisymmetric matrix $b_{\mu \nu },$ or general
quantized $\theta _{\mu \nu }$ 
\begin{equation}
\theta _{\mu \nu }=\frac{na^{2}}{4\pi }\left( b_{\mu \nu }\right) ^{-1}.
\end{equation}

\subsection{Map $\left( \hat{\Delta}\left( x\right) \right) _{J}^{J^{\prime
}}$ from position lattice to matrix}

Consider the Fourier transform of the matrix $\left( \hat{V}_{p}\right)
_{J}^{J^{\prime }}$ that represents all possible translations on the lattice 
\begin{eqnarray}
\left( \hat{\Delta}\left( x\right) \right) _{J}^{J^{\prime }} &\equiv
&\sum_{p\in \left( \tilde{T}_{n}\right) ^{d}}\left( \hat{V}_{p}\right)
_{J}^{J^{\prime }}\,\frac{e^{ip^{\mu }x_{\mu }}}{N},\quad x_{\mu }\in \left(
T_{n}\right) ^{d} \\
&=&\frac{1}{N}\sum_{p\in \left( \tilde{T}_{n}\right) ^{d}}\left( e^{ip^{\mu
}\left( x_{\mu }-X_{\mu }\right) }\right) _{J}^{J^{\prime }}\,.
\end{eqnarray}
The inverse transform is (recall $N=n^{d/2}$ or $n_{1}n_{2}\cdots n_{d/2}$) 
\begin{equation}
\left( \hat{V}_{p}\right) _{J}^{J^{\prime }}=\sum_{x_{\mu }\in \left(
T_{n}\right) ^{d}}\left( \hat{\Delta}\left( x\right) \right) _{J}^{J^{\prime
}}\,\frac{e^{-ip^{\mu }x_{\mu }}}{N},\quad p\in \left( \tilde{T}_{n}\right)
^{d}
\end{equation}
These finite Fourier transforms are defined with both positions and momenta
taken on lattices, and follow from the completeness and orthogonality
properties of the periodic lattice functions $f_{p}\left( x\right) $ 
\begin{equation}
f_{p}\left( x\right) =\frac{\exp \left( ip\cdot x\right) }{N},\quad p^{\mu
}\in \left( \tilde{T}_{n}\right) ^{d},\,\,x^{\mu }\in \left( T_{n}\right)
^{d},  \label{fourier}
\end{equation}
which are given by 
\begin{equation}
\sum_{p\in \left( \tilde{T}_{n}\right) ^{d}}\frac{e^{ip\cdot x}}{N}\frac{%
e^{-ip\cdot x^{\prime }}}{N}=\delta _{x,x^{\prime }}\,,\quad \sum_{x\in
\left( T_{n}\right) ^{d}}\frac{e^{ip\cdot x}}{N}\frac{e^{-ip^{\prime }\cdot
x}}{N}=\delta _{p,p^{\prime }}.\quad  \label{ortho}
\end{equation}
These are verified by performing finite sums, e.g. the sum over $p^{1}=-k_{2}%
\frac{2\pi b_{12}}{na}$ gives 
\begin{eqnarray}
&&\sum_{k_{2}=0}^{n-1}\frac{1}{n}e^{-ik_{2}\frac{2\pi b_{12}}{n}%
j_{1}}e^{ik_{2}\frac{2\pi b_{12}}{n}j_{1}^{\prime }} \\
&=&\sum_{k_{2}=0}^{n-1}\frac{1}{n}\left( \omega ^{j_{1}-j_{1}^{\prime
}}\right) ^{k_{2}}=\frac{1}{n}\frac{1-\left( \omega ^{j_{1}-j_{1}^{\prime
}}\right) ^{n}}{1-\omega ^{j_{1}-j_{1}^{\prime }}}=\delta _{j,j^{\prime }}.
\end{eqnarray}
The numerator is always zero since $\omega ^{n}=1$, but the denominator also
vanishes provided $j_{1}-j_{1}^{\prime }=0,$ thus $\delta _{j,j^{\prime }}$
is the correct answer. Likewise, in the definition of $\left( \hat{\Delta}%
\left( x\right) \right) _{J}^{J^{\prime }},$ by concentrating on any one of
the sums over $p^{\mu }$, e.g. $p^{1}=-k_{2}\frac{2\pi b_{12}}{n},$ using (%
\ref{Tp}, \ref{lk1k2}), one finds 
\begin{eqnarray}
&&\sum_{k_{2}=1}^{n-1}\hat{h}^{k_{2}}\omega _{12}^{k_{1}k_{2}/2}\,\,\exp
\left( i\,\frac{2\pi b_{12}}{n}j_{1}k_{2}\right) \\
&=&\sum_{k_{2}=1}^{n-1}\left( \hat{h}\left( \omega _{12}\,\right)
^{j_{1}+k_{1}/2}\right) ^{k_{2}}=\frac{1-\hat{h}^{n}\left( \left( \omega
_{12}\,\right) ^{n}\right) ^{j_{1}+k_{1}/2}}{1-\hat{h}\left( \omega
_{12}\,\right) ^{j_{1}+k_{1}/2}}.
\end{eqnarray}
The numerator is proportional to $1$ since $\hat{h}^{n}=1,$ and it vanishes
by using $\omega _{12}^{n}=1$ when $k_{1}$ is even (there is a further sum
over $k_{1})$. So, the result of the sum would be zero (for fixed even $%
k_{1})$ except for the fact that the matrix in the denominator also has one
eigenvalue that vanishes. In fact, formally $\left( \hat{\Delta}\left(
x\right) \right) _{J}^{J^{\prime }}$ are the matrix elements of the delta
function $\delta \left( x_{\mu }-X_{\mu }\right) ,$ with non-commutative
operators $X_{\mu },$ and lattice points $x_{\mu }\in \left( T_{n}\right)
^{d}.$

Under matrix multiplication $\hat{\Delta}\left( x\right) \hat{\Delta}\left(
y\right) $ satisfies a closed algebra, and yields a Kronecker delta function
upon tracing 
\begin{eqnarray}
\left( \hat{\Delta}\left( x\right) \hat{\Delta}\left( y\right) \right)
_{J}^{J^{\prime }} &=&\frac{1}{N}\sum_{z_{\mu }\in \left( T_{n}\right)
^{d}}\left( \hat{\Delta}\left( z\right) \right) _{J}^{J^{\prime
}}\,e^{2iB^{\mu \nu }\left( x_{\mu }-z_{\mu }\right) \left( y_{\nu }-z_{\nu
}\right) }\,  \label{product} \\
Tr\left( \hat{\Delta}\left( x\right) \hat{\Delta}\left( y\right) \right)
&=&N\,\delta _{x,y}  \label{trace} \\
Tr\left( \hat{\Delta}\left( x\right) \right) &=&N^{2}\,\delta _{x,0}.
\end{eqnarray}
These are derived by using the group algebra (\ref{galgeb}), thus 
\begin{eqnarray}
\hat{\Delta}\left( x\right) \hat{\Delta}\left( y\right) &=&\sum_{p,p^{\prime
}}\hat{V}_{p}\hat{V}_{p^{\prime }}\,\frac{e^{ip^{\mu }x_{\mu }}}{N}\,\frac{%
e^{ip^{\prime \mu }y_{\mu }}}{N},  \label{firstline} \\
&=&\sum_{p,p^{\prime }}\hat{V}_{p+p^{\prime }}\,\exp \left( -\frac{i}{2}%
\theta ^{\mu \nu }p_{\mu }p_{\nu }^{\prime }\right) \frac{e^{ip^{\mu }x_{\mu
}}}{N}\,\frac{e^{ip^{\prime \mu }y_{\mu }}}{N}, \\
&=&\sum_{p,p^{\prime }}\sum_{z}\hat{\Delta}\left( z\right) \frac{e^{-i\left(
p^{\mu }+p^{\prime \mu }\right) z_{\mu }}}{N}\exp \left( -\frac{i}{2}\theta
^{\mu \nu }p_{\mu }p_{\nu }^{\prime }\right) \frac{e^{ip^{\mu }x_{\mu }}}{N}%
\,\frac{e^{ip^{\prime \mu }y_{\mu }}}{N},  \nonumber
\end{eqnarray}
using the orthogonality/completeness relations (\ref{ortho}) to perform the
sum over $p_{\mu }$ one finds (\ref{product}). Also, using (\ref{trVp}) or (%
\ref{ortho}) in (\ref{firstline}) one derives (\ref{trace}).

\subsection{Diamond product}

We may now define functions on the lattice, such as gauge fields $A\left(
x\right) .$ Since there are only $N^{2}$ points on the lattice, these
functions really consist of only $N^{2}$ numbers. Therefore, it makes sense
to set up a map to a $N\times N$ matrix $\hat{A}_{J}^{J^{\prime }}$ by using
the map $\left( \hat{\Delta}\left( x\right) \right) _{J}^{J^{\prime }}$%
\begin{equation}
\hat{A}_{J}^{J^{\prime }}=\sum_{x_{\mu }\in \left( T_{n}\right) ^{d}}A\left(
x\right) \,\left( \hat{\Delta}\left( x\right) \right) _{J}^{J^{\prime
}},\quad A\left( x\right) =N^{-1}Tr\left( \hat{\Delta}\left( x\right) \hat{A}%
\right) .  \label{corresp}
\end{equation}
All the information in $A\left( x\right) $ on the $N^{2}$ lattice points is
contained in the $N^{2}$ entries of $\hat{A}_{J}^{J^{\prime }}.$ The matrix $%
\hat{A}$ can be viewed as an operator acting on the quantum Hilbert space of
non-commutative geometry. To define products among the $\hat{A}_{\mu }$ it
is natural to adopt the usual product of operators in quantum mechanics,
which in this case, corresponds to ordinary matrix product $\left( \hat{A}%
_{\mu }\hat{A}_{\nu }\right) _{J}^{J^{\prime }}.$ Having defined the
product, we introduce the diamond product in $x^{\mu }$ space as the one
that is equivalent to the matrix product via the map (\ref{corresp}) 
\begin{eqnarray}
\left( A_{\mu }\diamond A_{\nu }\right) \left( x\right) &\equiv
&N^{-1}Tr\left( \hat{\Delta}\left( x\right) \hat{A}_{\mu }\hat{A}_{\nu
}\right) ,  \label{diam} \\
\left( \hat{A}_{\mu }\hat{A}_{\nu }\right) _{J}^{J^{\prime }}
&=&\sum_{x_{\mu }\in \left( T_{n}\right) ^{d}}A_{\mu }\left( x\right)
\diamond A_{\nu }\left( x\right) \,\left( \hat{\Delta}\left( x\right)
\right) _{J}^{J^{\prime }}
\end{eqnarray}
This expression can be rewritten purely in terms of $A_{\mu }\left( x\right) 
$ by using the correspondence (\ref{corresp}) and then using the formulas in
(\ref{product}, \ref{trace}) 
\begin{eqnarray}
A_{\mu }\left( x\right) \diamond A_{\nu }\left( x\right)
&=&N^{-1}\sum_{y,z\in \left( T_{n}\right) ^{d}}Tr\left( \hat{\Delta}\left(
x\right) \hat{\Delta}\left( y\right) \hat{\Delta}\left( z\right) \right)
A_{\mu }\left( y\right) A_{\nu }\left( z\right)  \nonumber \\
&=&N^{-1}\sum_{y,z\in \left( T_{n}\right) ^{d}}e^{2iB^{\mu \nu }\left(
x_{\mu }-y_{\mu }\right) \left( x_{\nu }-z_{\nu }\right) }A_{\mu }\left(
y\right) A_{\nu }\left( z\right)  \label{diamond}
\end{eqnarray}
It is physically interesting to note that the sums in the diamond product
are weighted by exponentials of the flux that passes through the area
defined by the three lattice points $x,y,z$.

For the complete set of periodic functions $f_{p}\left( x\right) $ given in (%
\ref{fourier}) it is interesting to note their matrix map according to (\ref
{corresp}) 
\[
\left( \hat{f}_{p}\right) _{J}^{J^{\prime }}=\sum_{x_{\mu }\in \left(
T_{n}\right) ^{d}}\frac{\exp \left( ip\cdot x\right) }{N}\,\left( \hat{\Delta%
}\left( x\right) \right) _{J}^{J^{\prime }}=\left( \hat{V}_{-p}\right)
_{J}^{J^{\prime }} 
\]
The result of applying the diamond product on them is 
\begin{eqnarray}
\frac{\exp \left( ip\cdot x\right) }{N}\diamond \frac{\exp \left( ip^{\prime
}\cdot x\right) }{N} &=&\exp \left( -\frac{i}{2}\theta ^{\mu \nu }p_{\mu
}p_{\nu }^{\prime }\right) \frac{\exp \left( i\left( p+p^{\prime }\right)
\cdot x\right) }{N} \\
f_{p}\left( x\right) \diamond f_{p^{\prime }}\left( x\right) &=&\exp \left( -%
\frac{i}{2}\theta ^{\mu \nu }p_{\mu }p_{\nu }^{\prime }\right)
f_{p+p^{\prime }}\left( x\right) .
\end{eqnarray}
It is important to emphasize that all positions $x^{\mu }$ and all momenta $%
p^{\mu }$ are on their respective lattices with only $N^{2}$ allowed values
for each. The form of this result has a complete parallel in the continuum
limit when the positions and momenta are continuous and the star product (%
\ref{starp}) is used instead of the diamond product (\ref{diamond}) 
\begin{eqnarray}
e^{ip\cdot x}\ast e^{ip^{\prime }\cdot x} &=&\left. \exp \left( \frac{i}{2}%
\theta ^{\mu \nu }\partial _{\mu }^{x}\partial _{\nu }^{y}\right) e^{ip\cdot
x}e^{ip^{\prime }\cdot y}\right| _{y=x} \\
&=&\exp \left( -\frac{i}{2}\theta ^{\mu \nu }p_{\mu }p_{\nu }^{\prime
}\right) \,e^{i\left( p+p^{\prime }\right) \cdot x}.
\end{eqnarray}
Since the plane waves $e^{ip\cdot x}$ form a complete set of functions in
the continuum theory, this shows that the continuum limit of the diamond
product is the star product given in (\ref{starp}).

We have shown through (\ref{diam}) that the diamond product $\left( A_{\mu
}\diamond A_{\nu }\right) \left( x\right) $ is equivalent to the finite $%
N\times N$ matrix product $\left( \hat{A}_{\mu }\hat{A}_{\nu }\right)
_{J}^{J^{\prime }}$ thanks to the map $\,\left( \hat{\Delta}\left( x\right)
\right) _{J}^{J^{\prime }}.$ Going over to the continuum corresponds to a
particular large $N$ limit (there are many possible large $N$ limits since $%
N=n_{1}n_{2}\cdots n_{d/2},$ and any of the factors could be large in
independent ways). When all $n_{i}=n\rightarrow \infty $ are large, the star
product can be associated with the large $N$ limit of the diamond product.
Note that in taking the large $n$ limit to reach the continuum, one must
keep $B_{\mu \nu }=2\pi b_{\mu \nu }n/\left( na\right) ^{2}$ and the
products $an,$ and $nb_{\mu \nu }$ finite (the number of flux lines $b_{\mu
\nu }$ per lattice plaquette goes to zero as the lattice distance vanishes).

\section{Discrete non-commutative gauge theory}

To construct a gauge theory we also need to define a lattice version of the
derivative of fields, such as $\partial _{\mu }A_{\nu }\left( x\right) .$
When $x^{\mu }$ is on the lattice we will write symbolically $\hat{\partial}%
_{\mu }A_{\nu }\left( x\right) $ where $\hat{\partial}_{\mu }$ is a discrete
operation that we define. The simplest way is to define it in matrix space
as a commutator $-i\left[ \hat{P}_{\mu },\hat{A}_{\nu }\right]
_{J}^{J^{\prime }}$ with a fixed set of $N\times N$ matrices $\hat{P}_{\mu
}, $ and then map it to $x$-space using the map (\ref{corresp}) 
\begin{eqnarray}
\hat{\partial}_{\mu }A_{\nu }\left( x\right) &=&N^{-1}Tr\left( -i\left[ \hat{%
P}_{\mu },\hat{A}_{\nu }\right] \hat{\Delta}\left( x\right) \right)
\label{derlatt} \\
-i\left[ \hat{P}_{\mu },\hat{A}_{\nu }\right] _{J}^{J^{\prime }}
&=&\sum_{x_{\mu }\in \left( T_{n}\right) ^{d}}\hat{\partial}_{\mu }A_{\nu
}\left( x\right) \,\left( \hat{\Delta}\left( x\right) \right)
_{J}^{J^{\prime }}
\end{eqnarray}
The important property of the definition is that this lattice derivative is
distributive when the diamond product is used, 
\[
\hat{\partial}_{\mu }\left( A_{\nu }\left( x\right) \diamond A_{\lambda
}\left( x\right) \right) =\left( \hat{\partial}_{\mu }A_{\nu }\left(
x\right) \right) \diamond A_{\lambda }\left( x\right) +A_{\nu }\left(
x\right) \diamond \left( \hat{\partial}_{\mu }A_{\lambda }\left( x\right)
\right) 
\]
Furthermore, $\hat{\partial}_{\mu }$ is commutative $\hat{\partial}_{\mu }%
\hat{\partial}_{\nu }=\hat{\partial}_{\nu }\hat{\partial}_{\mu }$ if the
matrices $\hat{P}_{\mu }$ commute with each other $\hat{P}_{\mu }\hat{P}%
_{\nu }=\hat{P}_{\nu }\hat{P}_{\mu }.$ That is, one can immediately show 
\begin{equation}
\hat{\partial}_{\mu }\left( \hat{\partial}_{\nu }A_{\lambda }\left( x\right)
\right) =\hat{\partial}_{\nu }\left( \hat{\partial}_{\mu }A_{\lambda }\left(
x\right) \right) ,
\end{equation}
by using the definition (\ref{derlatt}) and matrix Jacobi identities.

\subsection{Non-commutative U(1) gauge theory}

With these definitions we give the covariant derivative applied on any
function $\psi \left( x\right) $ defined on the periodic lattice 
\begin{equation}
\hat{D}_{\mu }\psi \left( x\right) \equiv \hat{\partial}_{\mu }\psi \left(
x\right) -iA_{\mu }\left( x\right) \diamond \psi \left( x\right) +i\psi
\left( x\right) \diamond A_{\mu }\left( x\right)
\end{equation}
Both the function and the covariant derivative transform covariantly under
gauge transformations provided $A_{\mu }\left( x\right) $ also transforms as
follows 
\begin{eqnarray}
\delta A_{\mu }\left( x\right) &=&\hat{D}_{\mu }\Lambda \left( x\right) , \\
\delta \psi \left( x\right) &=&i\Lambda \left( x\right) \diamond \psi \left(
x\right) -i\psi \left( x\right) \diamond \Lambda \left( x\right) ,\quad \\
\delta \left( \hat{D}_{\mu }\psi \left( x\right) \right) &=&i\Lambda \left(
x\right) \diamond \left( \hat{D}_{\mu }\psi \left( x\right) \right) -i\left( 
\hat{D}_{\mu }\psi \left( x\right) \right) \diamond \Lambda \left( x\right)
\end{eqnarray}
Using the map $\left( \hat{\Delta}\left( x\right) \right) _{J}^{J^{\prime }}$
each one of these equations can be written in the equivalent matrix space
for $\hat{\psi},\hat{A}_{\mu }.$ The covariant derivative becomes the matrix
commutator 
\begin{equation}
\hat{D}_{\mu }\psi \left( x\right) \rightarrow \left[ -i\left( \hat{P}_{\mu
}+\hat{A}_{\mu }\right) ,\hat{\psi}\right]
\end{equation}
The transformation laws are 
\begin{equation}
\delta \hat{\psi}=i\left[ \hat{\Lambda},\hat{\psi}\right] ,\quad \delta \hat{%
A}_{\mu }=\left[ -i\left( \hat{P}_{\mu }+\hat{A}_{\mu }\right) ,\hat{\Lambda}%
\right]
\end{equation}
and the covariance can be checked explicitly by using matrix Jacobi
identities 
\[
\delta \left[ -i\left( \hat{P}_{\mu }+\hat{A}_{\mu }\right) ,\hat{\psi}%
\right] =i\left[ \hat{\Lambda},\left[ -i\left( \hat{P}_{\mu }+\hat{A}_{\mu
}\right) ,\hat{\psi}\right] \right] . 
\]

Next we can define the covariant field strength in position space and in
matrix space as the commutator of the covariant derivatives, with the usual
map relating the two 
\begin{eqnarray}
F_{\mu \nu }\left( x\right) &=&N^{-1}Tr\left( \hat{F}_{\mu \nu }\hat{\Delta}%
\left( x\right) \right) \\
\left( \hat{F}_{\mu \nu }\right) _{J}^{J^{\prime }} &=&\sum_{x_{\mu }\in
\left( T_{n}\right) ^{d}}F_{\mu \nu }\left( x\right) \,\left( \hat{\Delta}%
\left( x\right) \right) _{J}^{J^{\prime }}
\end{eqnarray}
and 
\begin{eqnarray}
F_{\mu \nu }\left( x\right) &=&-i\hat{D}_{[\mu }\diamond \hat{D}_{\nu ]}=%
\hat{\partial}_{[\mu }A_{\nu ]}\left( x\right) +A_{[\mu }\diamond A_{\nu
]}\left( x\right) \\
\left( \hat{F}_{\mu \nu }\right) _{J}^{J^{\prime }} &=&-i\left[ \left( \hat{P%
}_{\mu }+\hat{A}_{\mu }\right) ,\left( \hat{P}_{\nu }+\hat{A}_{\nu }\right) %
\right] _{J}^{J^{\prime }}=-i\left[ \hat{a}_{\mu },\hat{a}_{\nu }\right]
_{J}^{J^{\prime }}
\end{eqnarray}
where in the last line we have used the definition for the matrix $\hat{a}%
_{\mu }$ 
\begin{equation}
\left( \hat{a}_{\mu }\right) _{J}^{J^{\prime }}=\left( \hat{P}_{\mu }+\hat{A}%
_{\mu }\right) _{J}^{J^{\prime }}
\end{equation}
that appears everywhere. The matrices $\hat{P}_{\nu },\hat{A}_{\nu }$ appear
everywhere only in the combination $\hat{a}_{\mu }$, therefore the theory is
expressed only in terms of the matrix $\hat{a}_{\mu }$.

The action for the pure U$\left( 1\right) $ non-commutative gauge theory is
then written in either discrete position space or in matrix space 
\begin{eqnarray}
S &=&\frac{1}{4N^{2}}\sum_{x\in \left( T_{n}\right) ^{d}}F_{\mu \nu }\left(
x\right) \diamond F_{\mu \nu }\left( x\right) \\
&=&\frac{1}{4}\sum_{x\in \left( T_{n}\right) ^{d}}Tr\left( \hat{F}_{\mu \nu }%
\hat{F}_{\mu \nu }\right) =-\frac{1}{4}Tr\left[ \hat{a}_{\mu },\hat{a}_{\nu }%
\right] ^{2}.
\end{eqnarray}
To derive the last line from the first line one can use the map (\ref{diam})
for the product $F_{\mu \nu }\diamond F_{\mu \nu }$ and then use $%
N^{-3}\sum_{x\in \left( T_{n}\right) ^{d}}\hat{\Delta}\left( x\right) =\hat{1%
}.$

Matter, including fermions, can be added naturally both in the lattice and
the matrix formulation. The supersymmetric version is also straightforward.

\subsection{Non-commutative U$\left( M\right) $ gauge theory}

The U$\left( M\right) $ gauge theory is naturally constructed by attaching
indices on the gauge fields $\left( A_{\mu }\left( x\right) \right)
_{a}^{a^{\prime }}$ with $a,a^{\prime }=1,2,\cdots ,M.$ Then the diamond
product is combined with matrix product $\left( A_{\mu }\left( x\right)
\diamond A_{\mu }\left( x\right) \right) _{a}^{a^{\prime }}.$ In the matrix
version the matrix has the following indices $\hat{A}_{Ja}^{J^{\prime
}a^{\prime }}.$ This is equivalent to enlarging the direct product space $%
J=\left( j_{1}j_{3}\cdots j_{d-1}\right) $ to $\left( j_{1}j_{3}\cdots
j_{d-1}a\right) .$ From the point of view of our discussion we can interpret
the additional index $a$ as arising from two extra non-commuting dimensions $%
\left[ X_{d+1},X_{d+2}\right] =i\left( B^{-1}\right) _{d+1,d+2},$ with their
eigenvalues on the lattice $j_{d+1}\equiv a=1,2,\cdots ,n_{\left( d+2\right)
/2},$ where $n_{\left( d+2\right) /2}\equiv M.$ Then we can regard the U$%
\left( M\right) $ non-commutative gauge theory in $d$ dimensions, as a U$%
\left( 1\right) $ non-commutative gauge theory in $d+2$ dimensions. In the
matrix version its action takes the form 
\begin{equation}
S=-\frac{1}{4}Tr\left[ \hat{a}_{\mu },\hat{a}_{\nu }\right] ^{2}
\end{equation}
where now $\hat{a}_{\mu }$ is a $NM\times NM$ matrix given by 
\begin{equation}
\left( \hat{a}_{\mu }\right) _{Ja}^{J^{\prime }a^{\prime }}=\left( \hat{P}%
_{\mu }\right) _{J}^{J^{\prime }}\delta _{a}^{a^{\prime }}+\left( \hat{A}%
_{\mu }\right) _{Ja}^{J^{\prime }a^{\prime }}.
\end{equation}
Since $\left( \hat{A}_{\mu }\right) _{Ja}^{J^{\prime }a^{\prime }}$ is the
most general matrix, $\left( \hat{a}_{\mu }\right) _{Ja}^{J^{\prime
}a^{\prime }}$ is also the most general $NM\times NM$ matrix. The form $%
\left( \hat{P}_{\mu }\right) _{J}^{J^{\prime }}\delta _{a}^{a^{\prime }}$
that seems to be pulled out artificially serves only to distinguish between
the space directions and the internal directions.

If we take this point of view, the U$\left( 1\right) $ non-commutative gauge
field may be labelled by $A_{\mu }\left( x^{\mu },\vec{\sigma}\right) $
where $\sigma _{1},\sigma _{2}$ are the extra coordinates that take values
at the $M^{2}$ lattice points in the $\left( \sigma _{1},\sigma _{2}\right) $
plane. This point of view was explored a long time ago in \cite{bars}, where
it was shown that the U$\left( M\right) $ gauge transformations at finite $M$
may also be regarded as discrete diffeomorphism transformations of the
discrete torus. As discussed in \cite{bars} these discrete area preserving
transformations can be embedded in SL$\left( 2,Z_{M}\right) .$

The action above is not yet a full $d+2$ dimensional gauge theory because
two additional fields $A_{d+1}\left( x^{\mu },\vec{\sigma}\right) $ and $%
A_{d+2}\left( x^{\mu },\vec{\sigma}\right) $ (or their matrix counterparts $%
\hat{a}_{d+1}$ and $\hat{a}_{d+2}$ ) are missing. However, if the original U$%
\left( M\right) $ non-commutative gauge theory is enlarged by including two
additional scalars in the adjoint representation of $U\left( M\right) ,$
then those two scalars could be interpreted as the extra space components of
the gauge field in $d+2$ dimensions, to complete it to a full U$\left(
1\right) $ non-commutative gauge theory in $d+2$ dimensions.

As in the U$\left( 1\right) $ case, matter fields can be easily added and
the theory can be supersymmetrized.

\section{Outlook}

In this paper we have discussed a discrete version of non-commutative
geometry that arises in string theory in the $B$ field background. We have
presented a formalism that introduced the diamond product as a lattice
version of the star product, and thus suggested a cutoff version of
non-commutative gauge theory.

One may ask what relation could one establish between our results and some
other attempt at providing a non-commutative version of Wilson's lattice
gauge theory formalism. In the same way that non-commutative gauge theory in
the continuum can be recast as a usual gauge theory with an infinite number
of high derivative terms \cite{seibwitnonc}, we suspect that our results can
be rewritten as a complicated Wilsonian type lattice action. It would be
interesting to compare the 't Hooft limits of ordinary and non-commutative
Yang-Mills on the lattice and verify their equivalence as claimed in \cite
{suss} for the continuum.

The similarity to reduced models could be further explored. Wilson loop
variables for non-commutative Yang-Mills have their counterparts in the
reduced Yang-Mills theory, but now the tracing must be done over both
internal and external matrix indices $Ja$. It would be interesting to
understand the relevance of this formulation of Wilson loop variables in the
extrapolation of the $AdS/CFT$ correspondence in the presence of the
background $B$ field, as studied in \cite{nads}.

In our version one could analyze the theory at finite $N$ which provides a
cutoff. For a sufficiently small $N$ the analysis can be done with the help
of a computer. Also, since the action is very simple, analytic computations
may not be out of reach. \vspace{0.5cm}

{\bf Acknowledgments:} We thank O. Aharony, R. Corrado, E. Gimon, A.
Hashimoto, N. Itzhaki and E. Witten for discussions.


\begin{thebibliography}{99}
\bibitem{nonc}  A. Connes, M. R. Douglas and A. Schwarz, JHEP, {\bf 9802:003}
(1998); M. R. Douglas and C. Hull, JHEP {\bf 9802:008} (1998); V.
Schomerus, ``D-branes and Deformation Quantization",
hep-th/9903205; F. Ardalan, H. Arfaei and M.M. Sheikh-Jabbari, JHEP
9902 (1999) 016; for reviews and further references consult, M. R.
Douglas, hep-th/9901146.

\bibitem{seibwitnonc}  N. Seiberg and E. Witten,
``String Theory and Non-Commutative Geometry",
hep-th/9908142.

\bibitem{(7)}  J. Moyal, Proc. Camb. Phil. Soc. {\bf 45} (1949) 99.

\bibitem{(8)}  H. Weyl, {\it The Theory of Groups and Quantum Mechanics}
(Dover, N.Y. 1931).

\bibitem{(11)}  G. Baker, Phys. Rev. {\bf 109} (1958) 2198.

\bibitem{star}  F. Bayen, M. Flato, C. Fronsdal, A. Lichnerowicz and D.
Sternheimer, Ann. Phys. {\bf 111} (1978) 61-110 and 111-151.

\bibitem{bars}  I. Bars, USC-90/HEP-20,  unpublished
\hfill\break (KEK library http://ccdb1.kek.jp/cgi-bin/img\_index?9103241,
\hfill\break or see http://physics.usc.edu/\symbol{126}bars/papers/KEK.pdf
(or .ps or .dvi) );
H. Garcia-Compean, J.F. Plebanski and N. Quiroz-Perez, Int. Jour. Mod.
Phys. A13 (1998) 2089.

\bibitem{(1)}  J. Goldstone, unpublished; J. Hoppe, Int. J. Mod. Phys, {\bf %
A4} (1989) 5235; J. Hoppe, MIT Ph.D. Thesis, Elem. Part. Res. J. (Kyoto) 
{\bf 80} (89/90) no.3.

\bibitem{(2)}  D. B. Fairlie, P. Fletcher and C.K. Zachos, Phys. Lett. {\bf %
B218} (1989) 203. D.B. Fairlie and C.K. Zachos, Phys. Lett. {\bf B224}
(1990) 101; V. Arnold Ann. Inst. Fourier XVI, No.{\bf 1} (1966) 319.

\bibitem{(3)}  E. Floratos and J. Illiopoulos, Phys. Lett. {\bf B201} (1988)
237; E. Floratos, Phys. Lett. {\bf B228} (1989) 335. E. Floratos and S.
Nicolis, J. Phys. A31 (1998) 3961.

\bibitem{(6)}  I. Bars, Phys. Lett. {\bf B245} (1990) 35.

\bibitem{(4)}  C. N. Pope and L. Romans, Class. Quantum Grav. {\bf 7} (1990)
97; D. B. Fairlie and C. K. Zachos, J. Math. Phys. {\bf 31} (1990) 1088.

\bibitem{(10)}  I. Bars, C. Pope and E. Sezgin, Phys. Lett. {\bf B210}
(1988) 85.

\bibitem{mem}  B. de Wit, J. Hoppe and H. Nicolai, Nucl. Phys. {\bf B305}
(1988) 545.

\bibitem{cor} L. Cornalba and R. Schiappa, ``Matrix Theory Star Products
from the Born-Infeld Action", hep-th/9907211.

\bibitem{(9)}  G. 'tHooft, Comm. Math. Phys. {\bf 81} (1981) 267.

\bibitem{hoppe}  J. Hoppe, Phys. Lett. {\bf B215} (1988) 706; Projective
representations of cyclic groups have been studied in the mathematics
literature: see, for example A. O. Morris, J. London Math. Soc. (2) {\bf 7}
(1973) 235.

\bibitem{(12)}  T. Eguchi and H. Kawai, Phys. Rev. Lett. {\bf 48} (1982) 47;
G. Bhanot, U. Heller and H. Neuberger, Phys. Lett. {\bf B113} (1982) 47; 
{\bf B115} (1982) 237; G. Parisi, Phys. Lett. {\bf B112} (1982) 319; G.
Parisi and Z.Yi-Cheng, Phys. Lett. {\bf B114} (1982) 314; D. Gross and Y.
Kitazawa, Nucl. Phys. {\bf B206} (1982) 440; I. Bars, Phys. Lett. {\bf B116}
(1982) 57; I. Bars, M. G\"unaydin and S. Yankielowicz, Nucl. Phys. {\bf B219}
(1983) 81; V. A. Kazakov and A. A. Migdal, Phys. Lett. {\bf B116} (1982)
423; A. Gonzales-Arroyo and M. Okawa, Phys. Rev. {\bf D27} (1983) 2397.

\bibitem{ikkt}  H. Aoki, N. Ishibashi. S. Iso, H. Kawai, Y. Kitazawa and T.
Tada, ``Non-commutative Yang-Mills in IIB Matrix Model'', hep-th/9908141;
M. Li, Nucl. Phys. B499 (1997) 149. 

\bibitem{iikk}  N. Ishibashi. S. Iso, H. Kawai, Y. Kitazawa, ``Wilson Loops
in Non-commutative Yang Mills'', hep-th/9910004.

\bibitem{zak}  J. Zak, Phys. Rev. {\bf 134} (1964) A1602.

\bibitem{suss}  D. Bigatti and L. Susskind, ``Magnetic Fields, Branes, and
Non-Commutative Geometry",  hep-th/9908056; Z. Yin, ``A
note on Space Non-Commutativity", hep-th/9908152.

\bibitem{nads}  A. Hashimoto and N. Itzhaki,
``Non-Commutative Yang-Mills and the AdS/CFT
Correspondence", hep-th/9907166; J. Maldacena and J. Russo,
``Lanrge-N Limit of Non-Commutative Gauge Theories", hep-th/9908134.
\end{thebibliography}
\end{document}